\documentclass{Interspeech}

\interspeechcameraready

\title{WavShape: Information-Theoretic Speech Representation Learning\\ for Fair and Privacy-Aware Audio Processing}

\author[affiliation={1}*]{Oguzhan}{Baser}
\author[affiliation={1}*]{A. Ege}{Tanriverdi}
\author[affiliation={1,2}*]{Kaan}{Kale}
\author[affiliation={1}]{Sandeep}{Chinchali}
\author[affiliation={1,2}]{Sriram}{Vishwanath}

\affiliation{Department of Electrical and Computer Engineering}{The University of Texas at Austin}{USA}
\affiliation{Department of Electrical and Computer Engineering}{Georgia Institute of Technology}{USA}
\email{oguzhanbaser@utexas.edu}

\usepackage{comment}

\begin{document}

\maketitle
{\renewcommand{\thefootnote}{*}\footnotetext{The first three authors have equal contribution.}}

\begin{abstract}
Speech embeddings often retain sensitive attributes such as speaker identity, accent, or demographic information, posing risks in biased model training and privacy leakage. We propose WavShape, an information-theoretic speech representation learning framework that optimizes embeddings for fairness and privacy while preserving task-relevant information. We leverage mutual information (MI) estimation using the Donsker-Varadhan formulation to guide an MI-based encoder that systematically filters sensitive attributes while maintaining speech content essential for downstream tasks. Experimental results on three known datasets show that WavShape reduces MI between embeddings and sensitive attributes by up to 81\% while retaining 97\% of task-relevant information. By integrating information theory with self-supervised speech models, this work advances the development of fair, privacy-aware, and resource-efficient speech systems. 
\end{abstract}

\section{Introduction}

The speech processing field and audio-based machine learning (ML) has undergone a rapid transformation in recent years, driven by the advances in ML architectures such as transformers and self-supervised learning models \cite{baevski2020wav2vec, gulati20, whisper, elizalde2023clap}. These innovations have significantly enhanced speech recognition, speaker identification, and audio-based generative models which have several applications from real-time transcription to expressive text-to-speech synthesis and multilingual communication. However, despite these advancements, many challenges remain in fair and inclusive speech that generalizes across diverse populations, linguistic variations, and environmental conditions.  

A fundamental challenge in \textit{speech representation (SR) learning} is efficient extraction and compression of information from raw audio while preserving relevant features for downstream tasks. While ML models have achieved remarkable performance, their \textit{interpretability, fairness, and generalization} remain key concerns, particularly in real-world deployment scenarios where data imbalance, demographic bias, and privacy constraints must be accounted for \cite{leschanowsky24_spsc, feng2023review, swdbaser}. To address these concerns, an information-theoretic approach to \textit{speech encoding and representation learning} can provide a structured foundation for \textit{efficient, fair, and privacy-preserving} audio processing.  

Information theory (IT) has long served as the foundation of \textit{communication, signal processing, and data compression} \cite{shannon1948mathematical, cover1991elements}. It offers rigorous mathematical tools to characterize \textit{optimal representation learning}, minimizing redundancy while maximizing task-relevant information. In the context of speech processing, information-theoretic approaches have been applied to \textit{speech compression, feature selection, and speaker privacy} \cite{gray1980distortion, fleuret2004fast}. However, the full potential of these methods remains underexplored in \textit{modern deep learning-based speech models}. Specifically, the interplay between \textit{mutual information (MI)} and \textit{speech encoding} holds promise for developing \textit{task-aware, personalized, and fair SRs} that adapt to diverse speaker characteristics and linguistic variations.  

\begin{figure}
    \centering
    \includegraphics[width=\linewidth]{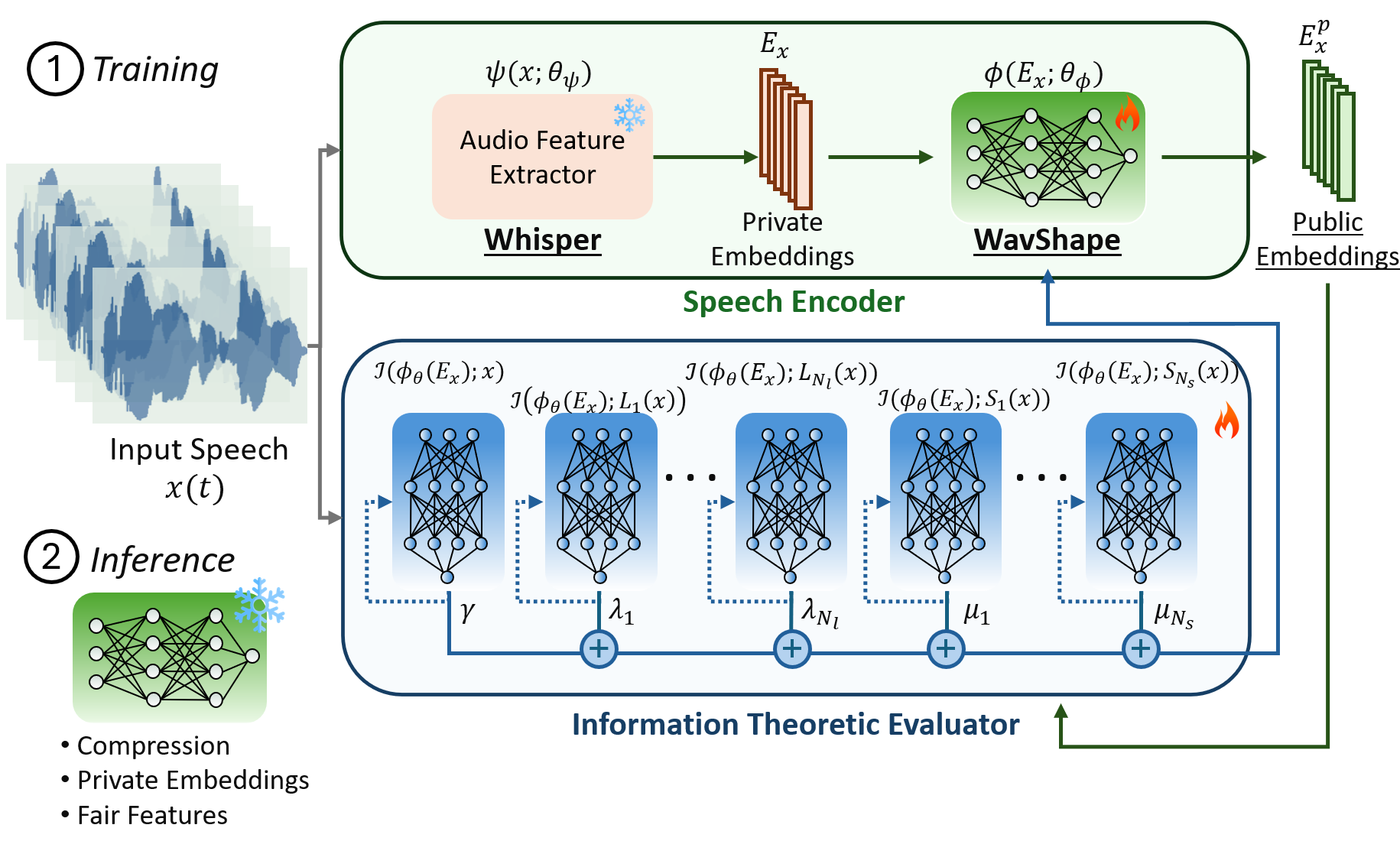}
    \caption{\textbf{How does WavShape encode fair and bias-free speech representations?} Our framework first extracts speech embeddings using a pre-trained speech encoder (e.g., Whisper). These embeddings, which may contain private and biased attributes, are then transformed via the WavShape projection layer to generate bias-mitigated public embeddings. The IT evaluator (blue) optimizes WavShape by minimizing the MI between speech embeddings and sensitive attributes while maximizing MI with task-relevant features and original representation. During training, the feature extractor is frozen, only WavShape and the IT evaluator are updated. At inference, the IT evaluator is removed, and WavShape is used in conjunction with the base feature extractor (green), ensuring bias-free and privacy-preserving embeddings for fair and inclusive modeling.\label{fig:architecture}}
\end{figure}
This paper represents an early step toward bridging IT principles with \textit{SR learning}, particularly in self-supervised and transformer-based models. We introduce an \textit{optimal adaptive audio encoding framework} that integrates MI-based learning objectives to \textit{ improve fairness, privacy, and efficiency of SRs}. Building upon our work in text domain, our approach extends \textit{MI maximization and variational inference} to \textit{speech feature learning}, enabling a principled mechanism for \textit{interpretable and bias-aware compression of speech signals}.\\
\noindent\textit{\underline{Literature Review:}} Despite recent advancements in ASR and SR learning, the integration of MI remains underexplored. Existing MI-based methods in ML focus on enhancing the SR for performance \cite{ispmi1,ispmi2,ispmi3,ispmi4,ispmi5}, but none systematically address fairness and privacy constraints in speech processing. Conversely, fairness- and security-aware works exist \cite{ispfairness1, ispfairness2, ispfairness3, securespectra} but lack a mathematical framework to incorporate these objectives into training. Recent works in other domains have successfully leveraged MI-based optimization to achieve privacy-preserving and fairness-aware encoding \cite{esfahanizadeh2023infoshape, kale2024texshape}, highlighting an urgent need for MI-driven approaches that ensure fair, privacy-secure, and efficient SRs. This work fills the gap by introducing a MI-based framework for structured, fairness-aware, and privacy-preserving speech processing. In the light of prior work, our contributions are threefold:

\begin{itemize}
    \item We opensource\footnote{\url{www.github.com/UTAustin-SwarmLab/WavShape}} an information-theoretic architecture for fair and private speech processing, leveraging Whisper \cite{whisper} and variational MI inference \cite{donsker1975asymptotic, kraskov2004estimating}.  
    \item We introduce MI-based encoding techniques to optimize SRs for \textit{fairness, privacy, and efficient resource allocation}, ensuring that compressed speech features retain maximal task-relevant information.  
    \item We explore how MI estimation techniques can serve as an alternative to traditional \textit{distortion metrics} in speech compression, leading to \textit{more interpretable, fair, and adaptable speech models} in real-world settings.
\end{itemize}

By integrating MI into \textit{speech processing ML architectures}, this work aims to advance \textit{inclusive and equitable speech technologies}, contributing to \textit{robust, privacy-preserving, and resource-efficient AI systems} that serve diverse user populations.

\section{Methodology}
Here, we detail our architecture and its theoretical foundations. 
\subsection{MI in Speech Representation Learning}
Consider two random variables, $\mathbf{X} \in \mathcal{X}$ and $\mathbf{E} \in \mathcal{E}$, where $\mathcal{X}$ represents the space of \textit{raw speech spectogram}, and $\mathcal{E}$ denotes the space of \textit{compressed or transformed SRs}. The MI between $\mathbf{X}$ and $\mathbf{E}$, denoted as $\mathcal{I}(\mathbf{X};\mathbf{E})$, quantifies how much information about $\mathbf{X}$ is preserved when mapped to $\mathbf{E}$. Following Shannon’s definition \cite{shannon1948mathematical}, MI is:
\begin{equation}
    \mathcal{I}(\mathbf{X};\mathbf{E}) = \sum_{X \in \mathcal{X}} \sum_{E \in \mathcal{E}} P_{\mathbf{X, E}}(X,E) \log \left(\frac{P_{\mathbf{X,E}}(X,E)}{P_{\mathbf{X}}(X) P_{\mathbf{E}}(E)}\right),
\end{equation}
where $P_{\mathbf{X,E}}(X,E)$ is the \textit{joint probability distribution} of the original and transformed SRs, while $P_{\mathbf{X}}(X)$ and $P_{\mathbf{E}}(E)$ are the \textit{marginal probability distributions}. When dealing with \textit{continuous speech signals}, summations are replaced with integrals, and mass functions are replaced with density functions.

In the context of \textit{SR learning}, maximizing MI ensures that essential \textit{acoustic, linguistic, and paralinguistic} features are preserved in the transformed representation while filtering out redundant or sensitive information. This is particularly critical in designing \textit{fair and privacy-aware speech processing models} that generalize across diverse speaker populations. 

A practical MI formulation based on the \textit{Donsker-Varadhan representation} of the Kullback-Leibler divergence \cite{donsker1975asymptotic} is:
\begin{equation}
    \mathcal{I}(\mathbf{X};\mathbf{E}) = \sup_{F: \Omega \rightarrow \mathbb{R}} \mathbb{E}_{P_{\mathbf{X,E}}} [F(\mathbf{X},\mathbf{E})] - \log \mathbb{E}_{P_{\mathbf{X}} P_{\mathbf{E}}} [e^{F(\mathbf{X},\mathbf{E})}].\label{eqn:approxMI}
\end{equation}

Here, the function $F(\mathbf{X}, \mathbf{E})$ acts as a discriminator that approximates MI. This optimization can be solved using ML, where $F$ is parameterized as a neural network, trained via SGD \cite{sgd}. In practice, MI estimation is achieved by replacing the expectations with empirical averages computed over mini-batches sampled from the joint and marginal distributions.

\subsection{Optimizing The Speech Representation for Fairness}
Our objective is to design a \textit{learned encoder} $\psi_{\Theta_\psi},\phi_{\Theta_\phi}: \mathcal{X} \to \mathcal{E}$ that transforms raw speech signals $\mathbf{X}$ into compressed and structured embeddings $\mathbf{E}$, while ensuring fairness, efficiency, and privacy. The encoder is optimized as follows:
\begin{equation}
    \begin{aligned}
        \max_{\Theta_\phi} & \, \gamma \mathcal{I}(\phi_{\Theta_\phi}(\mathbf{X});\mathbf{X}) + \sum_{i=1}^{N_l} \lambda_i \mathcal{I}(\phi_{\Theta_\phi}(\mathbf{X});T_i(\mathbf{X})) \\
        & - \sum_{j=1}^{N_s} \mu_j \mathcal{I}(\phi_{\Theta_\phi}(\mathbf{X});S_j(\mathbf{X})),
    \end{aligned}
\end{equation}
where $\Theta_\phi$ are the trainable parameters of the encoder, $T_i(\mathbf{X})$ represents the \textit{i-th task-relevant acoustic or linguistic feature} (e.g., phonetic content, speaker identity, or prosody), $S_j(\mathbf{X})$ represents the \textit{j-th sensitive attribute} (e.g., speaker demographics, accent, or background noise characteristics), $\gamma, \lambda_i$, and $\mu_j$ are tunable hyperparameters that balance the trade-offs between information preservation, utility, and privacy. The dimensionality of $\mathbf{E}$ controls the level of compression applied to the SR.

Unlike traditional approaches that rely on heuristic distortion metrics (e.g., mean squared error in reconstructed waveforms), this formulation explicitly models task-relevant and sensitive information, allowing SRs to be optimized for multiple privacy and fairness-aware objectives. For simplicity, we henceforth denote $\Theta_\phi$ as $\Theta$ since it is the trainable part in $\psi_{\Theta_\psi},\phi_{\Theta_\phi}$. 

The optimization objective consists of three key terms, each with a separate physical interpretation. (1) \textit{Information-Preserving Compression} ($\gamma \mathcal{I}(\phi_\Theta(\mathbf{X});\mathbf{X})$) ensures that essential information from the speech is retained with less redundancy. Unlike classical compressions minimizing reconstruction loss, this approach explicitly maximizes MI, leading to task-specific compression that preserves relevant characteristics. (2) \textit{Task-Oriented Speech Encoding} ($\sum_{i=1}^{N_l} \lambda_i \mathcal{I}(\phi_\Theta(\mathbf{X});T_i(\mathbf{X}))$) enhances representations to be more informative for specific downstream tasks (e.g., speech recognition, emotion detection, speaker verification). It is useful for scenarios where selective preservation of acoustic cues is crucial, such as emotion recognition in noisy environments or speaker adaptation in multilingual systems. (3) \textit{Fairness and Privacy Constraints} ($-\sum_{j=1}^{N_s} \mu_j \mathcal{I}(\phi_\Theta(\mathbf{X});S_j(\mathbf{X}))$) suppresses any unwanted information that could introduce \textit{biases or privacy risks}, such as speaker gender, accent, or ethnicity. Hence, the models trained on encoded speech embeddings do not unfairly discriminate against certain demographics.

We offer a principled approach to learning task-optimized and fairness-aware SRs. Specifically, it can prevent unintended biases in ASR by filtering out accent and demographic-related features or embeddings (e.g., age, gender), minimize personally identifiable information while preserving linguistic content.

\subsection{WavShape Architecture}
Following the theory, we present \texttt{WavShape}, an information-theoretic representation learning framework for speech. It consists of two components as shown in Fig. \ref{fig:architecture}. \textit{A speech encoder} (in green) transforms raw audio into structured representations while preserving task-relevant and fairness-aware speech properties. \textit{An information-theoretic evaluator} (in blue) acts as a differentiable loss function to guide the encoder’s training.  

The trained encoder can be deployed across diverse speech datasets, making it applicable to bandwidth-efficient speech transmission, privacy-preserving voice communication, and fairness-aware speech modeling. The computational cost of using \texttt{WavShape} is comparable to the inference cost of self-supervised or transformer-based speech models.

\noindent\textit{\underline{Speech Encoder}} consists of a two-stage transformation. (1) \textit{Feature Extraction Layer} (Non-trainable) uses a pre-trained embedding model (e.g., wav2vec 2.0 \cite{baevski2020wav2vec}, CLAP \cite{elizalde2023clap}, Whisper \cite{whisper}) to convert raw waveforms into an intermediate representation that encodes phonetic, prosodic, and linguistic information. This ensures robustness to speaker variability and environmental noise while maintaining essential acoustic properties. (2) \textit{Information-Theoretic Embedding Layer} (Trainable) maps the extracted features into lower-dimensional speech embeddings. We optimize this to preserve task-relevant information while filtering out sensitive or redundant attributes (e.g., speaker identity, accent), guided by MI-based constraints described previously, ensuring that encoded representations balance efficiency, fairness, and privacy. (1) and (2) together enable task-adaptive speech encoding, where the embedding retains maximal useful information while suppressing  bias or privacy leaks.

\noindent\textit{\underline{Information-Theoretic Evaluator:}} plays a critical role in optimizing the encoder. It estimates the MI terms required for training by evaluating the information retained between raw and encoded SRs $\mathcal{I}(\mathbf{X};\mathbf{E})$, the utility of the representation for specific tasks $\mathcal{I}(\mathbf{E}; T_i(\mathbf{X}))$, and the removal of sensitive attributes $\mathcal{I}(\mathbf{E}; S_j(\mathbf{X}))$. Following prior work on MI estimation \cite{belghazi2018mine, songunderstanding}, we adopt the Donsker-Varadhan MI formulation, which allows MI to be approximated using a discriminator network trained via SGD as in Eq. \ref{eqn:approxMI}. By using universal approximator ML models \cite{hornik1989multilayer} to approximate MI, the evaluator efficiently quantifies how well the speech encoding preserves desired properties while discarding unwanted features.

\subsection{Training Procedure}

Training involves a dual optimization where the encoder and the MI evaluator are updated iteratively.

\noindent\textit{\underline{Encoder Training (Epoch-Level Optimization):}} At each epoch, the encoder weights $\Theta$ are updated based on the MI-based loss function. The optimizer maximizes task-relevant information retention while minimizing sensitive attribute leakage. The encoded representations are progressively refined to optimize fairness, privacy, and efficiency constraints.

\noindent\textit{\underline{MI Estimation (Iteration-Level Optimization):}} At each iteration within an epoch, we estimate the MI terms using a batch of speech samples \textcolor{black}{and update it in each iteration}. The MI estimator refines its predictions over time, guiding better the encoder’s training. As training progresses, the number of MI estimation iterations required per epoch decreases for faster convergence.

\noindent After sufficient training, we use the encoder separately for SR learning, compression, and privacy-aware speech analytics.

\begin{figure}
    \centering
    \includegraphics[width=0.7\linewidth]{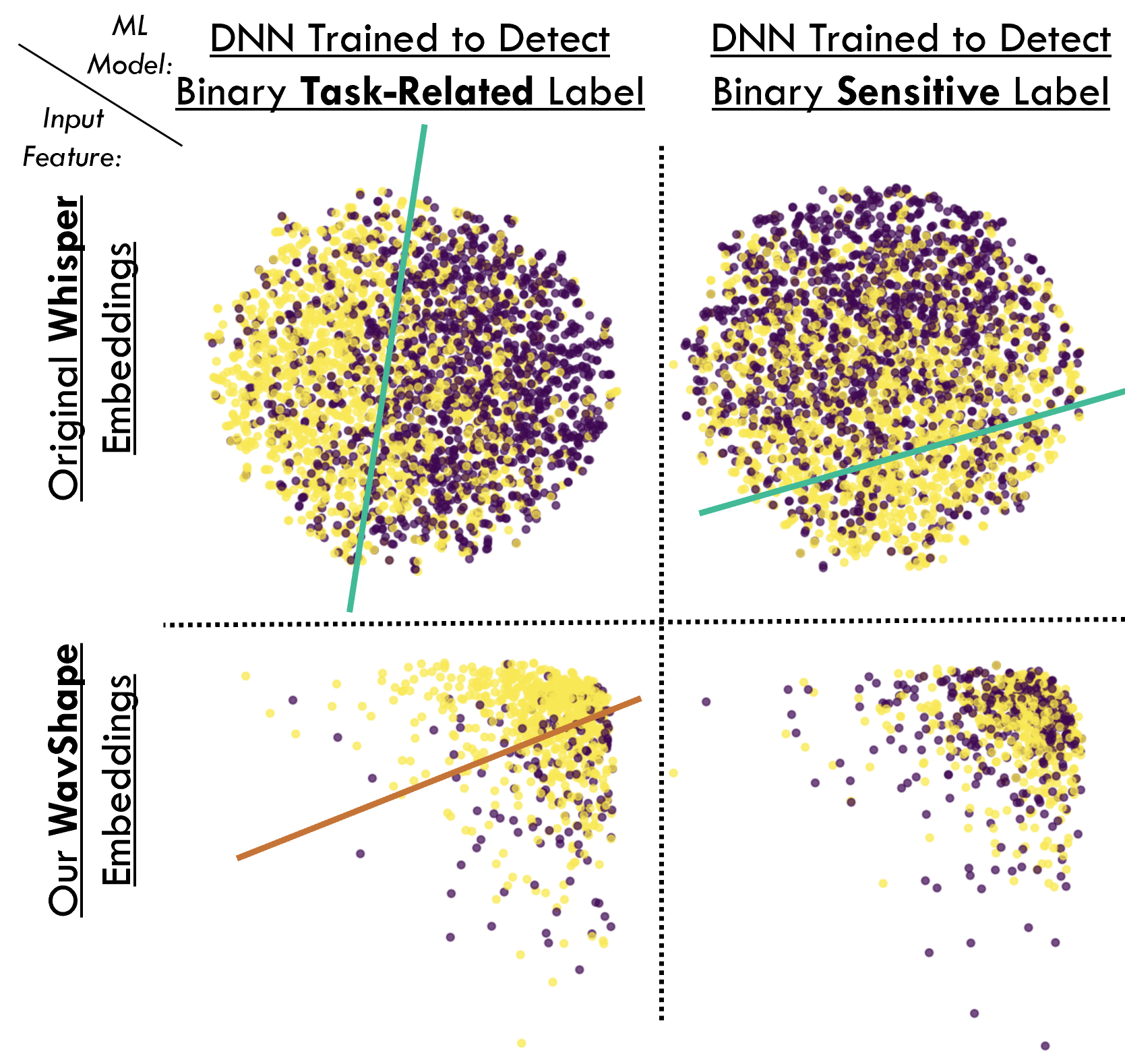}
    \caption{\small \textcolor{black}{We train two pairs of classifiers to predict task-relevant (left) and sensitive (right) binary labels (yellow and purple). We train one pair with Whisper embeddings while the other gets our embeddings as input. We project the whole CV dataset into a 2D t-SNE space with the trained classifiers and visualize the best decision boundary.  The task-relevant structure remains intact while sensitive attributes are effectively obscured compared to the models trained with original embeddings.} \label{fig:tsne}}
\end{figure}

\subsection{Use Cases}

The encoded SRs produced by \texttt{WavShape} can be deployed in several real-world applications. Voice assistant models perform better for standard English accents, disadvantaging non-native speakers. WS reduces the bias by minimizing MI between accents and embeddings, improving fairness. In telehealth, it suppresses health-revealing cues to protect privacy.

\noindent\textit{\underline{Efficient Speech Transmission (Compression)}}: It reduces bandwidth by encoding speech into low-dimensional representations that retain essential phonetic and prosodic features, useful for low-latency streamingor real-time speech synthesis.

\noindent \textit{\underline{Privacy Preserving Models (Anonymization)}}: WS filters out biometric or demographic attributes to anonymize interactions. 

\noindent \textit{\underline{Bias-Resistant Speech Recognition (Fairness)}}: It suppresses sensitive speaker attributes (e.g., gender, regional dialect) to reduce bias for more inclusive and equitable voice systems.

\noindent WS is \textbf{model-agnostic}. It can project any ASR encoder output into a compressed, bias-controlled space after training. The ASR's decoder can be fine-tuned or used solely as WS loss.

\section{Experimental Setup}
Here, we present the datasets, encoders, baselines, and metrics.\\
\noindent\underline{\textbf{Datasets:}} We evaluate our MI-driven speech encoding framework on three datasets. The \textbf{Mozilla Common Voice} (CV) dataset \cite{mozilladata} provides speech recordings with metadata on \textbf{age, gender, and accent}. We encode \textbf{age as the task label} while suppressing \textbf{gender and accent}, ensuring privacy-preserving and bias-mitigated SRs for fair ASR.
The \textbf{VCTK Corpus} \cite{veaux2017cstr} comprises recordings from 109 English speakers representing a diverse range of accents. Each participant reads around 400 sentences. Its labels have each speaker’s gender, age and region.\\
\noindent \underline{\textbf{Models:}} We utilize \textbf{Whisper} \cite{whisper} to extract features and \textbf{lightweight neural networks} for classification. Whisper is a transformer-based ASR model trained on 680,000 hours of multilingual speech, making it highly robust to acoustic variability and noise. Its self-supervised representations capture rich phonetic, linguistic, and contextual features, making it an ideal front-end encoder for our framework. By extracting its embeddings, we ensure task-relevant information is retained while enabling MI-driven filtering of sensitive attributes.  We train lightweight DNN on MI-optimized embeddings to quantify privacy leakage, fairness constraints, and task performance. \\
\noindent\underline{\textbf{Baselines:}}
We compare four types of embeddings. \textbf{Original} is the raw Whisper embeddings as the upper bound for utility and leakage. \textbf{Random} is the outputs from a randomly initialized encoder to assess informativeness without learning. \textbf{Noisy} is the Whisper embeddings with Gaussian differential privacy (DP) noise \cite{dwork2006differential}. \textbf{Encoded} (Ours) is the WavShape-optimized embeddings that explicitly retain task-relevant and suppress sensitive attributes. \textcolor{black}{We exclude adversarial training as it lacks control over removed information unlike these baselines.}

\noindent\underline{\textbf{Evaluation Metrics:}} \textbf{t-SNE} \cite{tsne} projects the embeddings into 2D space, revealing how task-related and sensitive attributes are encoded. Comparing pre- and post-encoding t-SNE plots, we validate that WS preserves task-relevant structure while suppressing sensitive information. \textbf{AUROC} \cite{auroc} measures a classifier’s ability to distinguish between classes. High AUROC for task-relevant and low AUROC for sensitive labels confirm that WS retains useful features while reducing private attribute leakage. \textcolor{black}{Also, we study \textbf{MI} as a metric to see how privacy (sensitive) and utility (task-related) evolve over epochs.} 

\noindent\underline{\textbf{Hyperparameters:}} \textcolor{black}{We detail the architecture and training configurations at the project's repository due to the limited space.} 

\begin{figure}[!t]
    \centering
    \includegraphics[width=\linewidth]{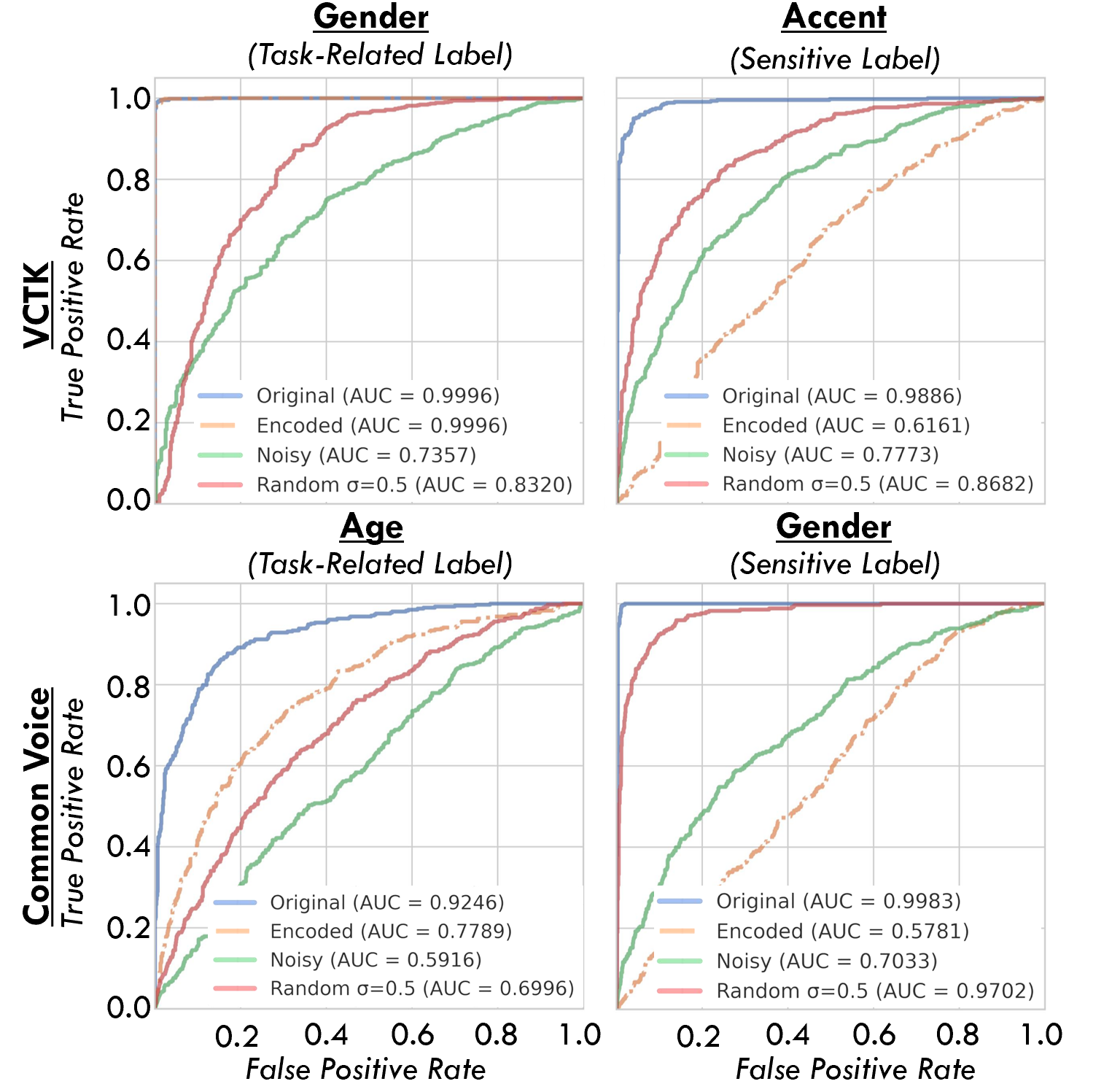}
    \caption{\small ROC and AUROC scores for the classifiers trained to predict task-relevant attributes (left) and sensitive attributes (right). The results demonstrate that our MI-based encoding preserves task-relevant distinctions while effectively reducing the predictability of sensitive attributes.\label{fig:rocs}}
\end{figure}

\section{Results}
\begin{figure}[!t] 
    \centering
    \includegraphics[width=\linewidth]{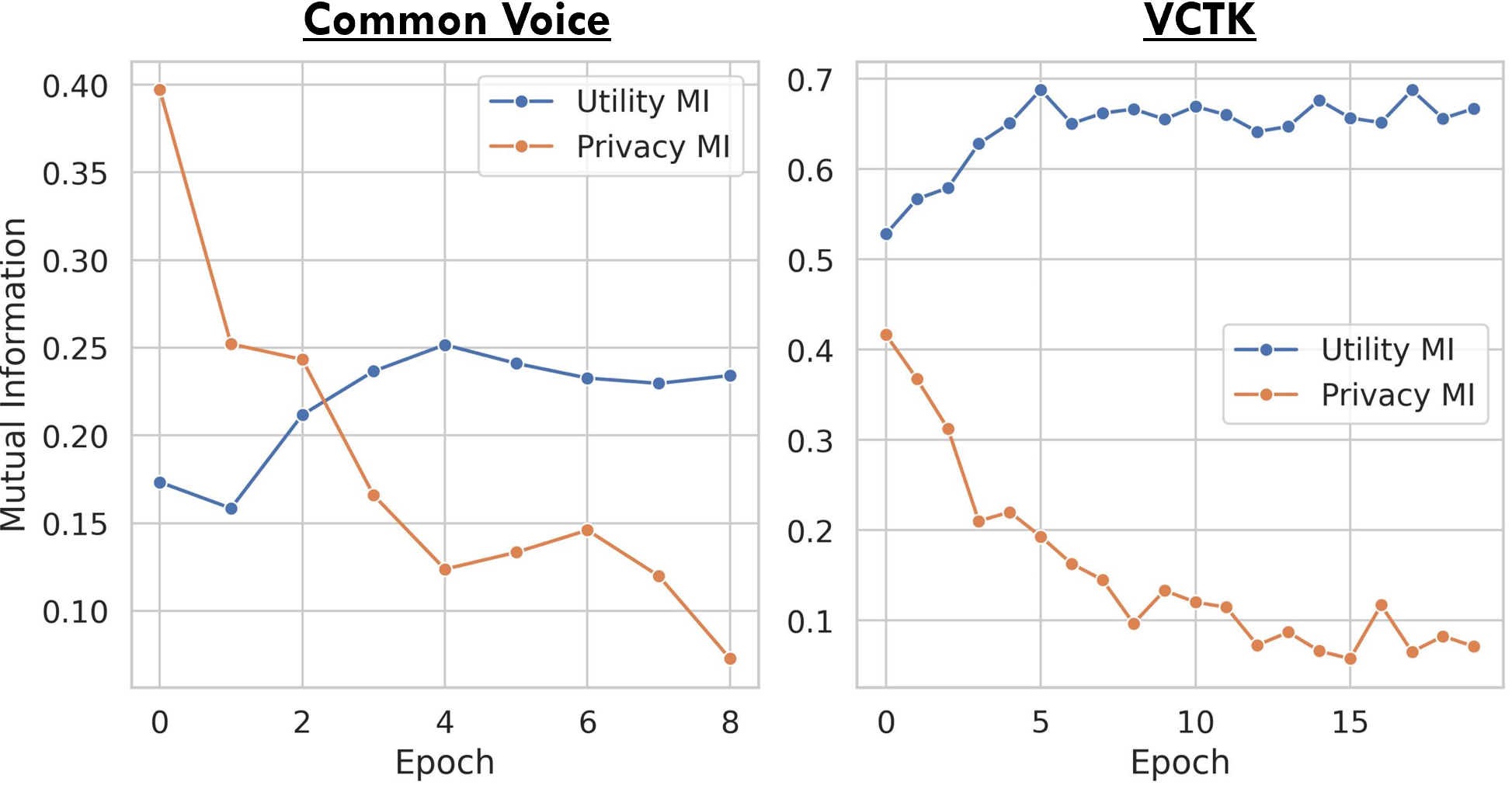}
    \caption{\textcolor{black}{IT Estimator's mean MI values for task-related $T_i(.)$ (utility) and sensitive $S_i(.)$ (privacy) features over the number of epochs. WS embeddings get more descriptive for the tasks while suppressing sensitive information over time.}}
    \label{fig:MI-res}
\end{figure}

WS produces compressed embeddings for privacy-preserving and fair SR. We validate our approach using two speech datasets, aiming to compress representations while preserving task-relevant information and filtering out sensitive attributes. The optimization objective simplifies to: $\max_\Theta  \mathcal{I}(\phi_\Theta(\mathbf{X}); T(\mathbf{X})) - \mu \mathcal{I}(\phi_\Theta(\mathbf{X}); S(\mathbf{X})),$
where $T(\mathbf{X})$ represents age in CV dataset and gender in VCTK dataset as target labels, while $S(\mathbf{X})$ denotes accent in VCTK dataset and gender in CV dataset as sensitive labels. WS compresses the embeddings from \textbf{512 to 64} dimensions and uses 1024 batch, $90\mathsf{x}$ stable for MI estimation as DV error scales linearly with the dimension and inversely with the batch size.. We compute MI between the WS embeddings and labels. MI estimation is performed over mini-batches spanning the full dataset, ensuring stability and robustness. To qualitatively assess the privacy and fairness, we projected the embeddings before and after WS transformation using t-SNE as in Fig. \ref{fig:tsne}. The MI results in Fig. \ref{fig:MI-res} show that the encoded representations in CV dataset \textit{retain separability for task-relevant information while making sensitive attributes indistinguishable}, as evidenced by a reduction in \textbf{the MI between embeddings and sensitive label ``gender'' from 0.40029 to 0.07493}, while \textbf{task-relevant MI (between embeddings and ``age'') is maintained at 0.23569}. Similarly, for the VCTK dataset, \textbf{the MI between embeddings and sensitive label ``accent'' decreases from  0.41647 to 0.07114 while the MI between embeddings and target label ``gender'' remained around 0.65510}. The ROC curves in Fig. \ref{fig:rocs} reveal that the task-relevant area under the ROC curves (AUROC) remains comparable to the original embeddings for both datasets, while there is a \textbf{38.36\%} and a \textbf{25.77\%}  decrease in the sensitive attribute AUROC for VCTK and CV respectively, confirming that our representation significantly reduces \textbf{privacy leakage and bias propagation} without compromising task utility.

\section{Conclusion}

This work integrates MI with SR learning, offering a principled approach for task-adaptive, privacy-preserving, and fairness-aware speech processing. By optimizing speech embeddings through IT constraints, we demonstrate how learned representations can balance utility, fairness, privacy, and resource efficiency, addressing the challenges in lossy compression, unbiased speech modeling, and privacy-aware voice data sharing. 

Our framework establishes a structured mechanism for controlling speech feature retention and suppression, leveraging a weighted optimization objective that enables fine-grained trade-offs between competing goals. While our current formulation employs a linear combination of MI-based objectives, future work will explore nonlinear optimization strategies that dynamically adjust trade-offs through reinforcement learning and multi-objective optimization techniques. Another key direction is the development of automated hyperparameter tuning mechanisms, informed by rate-distortion theory and DP, to optimize fairness and privacy constraints without manual selection.

Ultimately, this work lays the foundation for bias-resistant, privacy-secure, and resource-efficient speech models, contributing to the broader goal of fair and inclusive speech technology.
\newpage
\section{Acknowledgments}
This work was supported in part by the National Science Foundation grants under No. 2148186 and No. 2133481. Any opinions, findings, and conclusions or recommendations expressed in this material are those of the authors and do not necessarily reflect the views of the National Science Foundation.

\bibliographystyle{IEEEtran}
\bibliography{mybib}

\begin{thebibliography}{10}
\providecommand{\url}[1]{#1}
\csname url@samestyle\endcsname
\providecommand{\newblock}{\relax}
\providecommand{\bibinfo}[2]{#2}
\providecommand{\BIBentrySTDinterwordspacing}{\spaceskip=0pt\relax}
\providecommand{\BIBentryALTinterwordstretchfactor}{4}
\providecommand{\BIBentryALTinterwordspacing}{\spaceskip=\fontdimen2\font plus
\BIBentryALTinterwordstretchfactor\fontdimen3\font minus \fontdimen4\font\relax}
\providecommand{\BIBforeignlanguage}[2]{{%
\expandafter\ifx\csname l@#1\endcsname\relax
\typeout{** WARNING: IEEEtran.bst: No hyphenation pattern has been}%
\typeout{** loaded for the language `#1'. Using the pattern for}%
\typeout{** the default language instead.}%
\else
\language=\csname l@#1\endcsname
\fi
#2}}
\providecommand{\BIBdecl}{\relax}
\BIBdecl

\bibitem{baevski2020wav2vec}
A.~Baevski, Y.~Zhou, A.~Mohamed, and M.~Auli, ``wav2vec 2.0: A framework for self-supervised learning of speech representations,'' \emph{Advances in neural information processing systems}, vol.~33, pp. 12\,449--12\,460, 2020.

\bibitem{gulati20}
A.~Gulati, J.~Qin, C.-C. Chiu, N.~Parmar, Y.~Zhang, J.~Yu, W.~Han, S.~Wang, Z.~Zhang, Y.~Wu, and R.~Pang, ``Conformer: Convolution-augmented transformer for speech recognition,'' in \emph{Interspeech 2020}, 2020, pp. 5036--5040.

\bibitem{whisper}
A.~Radford, J.~W. Kim, T.~Xu, G.~Brockman, C.~McLeavey, and I.~Sutskever, ``Robust speech recognition via large-scale weak supervision,'' in \emph{International conference on machine learning}.\hskip 1em plus 0.5em minus 0.4em\relax PMLR, 2023, pp. 28\,492--28\,518.

\bibitem{elizalde2023clap}
B.~Elizalde, S.~Deshmukh, M.~Al~Ismail, and H.~Wang, ``Clap learning audio concepts from natural language supervision,'' in \emph{ICASSP 2023-2023 IEEE International Conference on Acoustics, Speech and Signal Processing (ICASSP)}.\hskip 1em plus 0.5em minus 0.4em\relax IEEE, 2023, pp. 1--5.

\bibitem{leschanowsky24_spsc}
A.~Leschanowsky and S.~Das, ``Examining the interplay between privacy and fairness for speech processing: A review and perspective,'' in \emph{4th Symposium on Security and Privacy in Speech Communication}, 2024, pp. 1--11.

\bibitem{feng2023review}
T.~Feng, R.~Hebbar, N.~Mehlman, X.~Shi, A.~Kommineni, S.~Narayanan \emph{et~al.}, ``A review of speech-centric trustworthy machine learning: Privacy, safety, and fairness,'' \emph{APSIPA Transactions on Signal and Information Processing}, vol.~12, no.~3, 2023.

\bibitem{swdbaser}
O.~Baser, M.~Yavuz, K.~Ugurlu, F.~Onat, and B.~U. Demirel, ``Automatic detection of the spike-and-wave discharges in absence epilepsy for humans and rats using deep learning,'' \emph{Biomedical Signal Processing and Control}, vol.~76, p. 103726, 2022.

\bibitem{shannon1948mathematical}
C.~E. Shannon, ``A mathematical theory of communication,'' \emph{The Bell system technical journal}, vol.~27, no.~3, pp. 379--423, 1948.

\bibitem{cover1991elements}
T.~M. Cover, \emph{Elements of information theory}.\hskip 1em plus 0.5em minus 0.4em\relax John Wiley \& Sons, 1999.

\bibitem{gray1980distortion}
R.~Gray, A.~Buzo, A.~Gray, and Y.~Matsuyama, ``Distortion measures for speech processing,'' \emph{IEEE Transactions on Acoustics, Speech, and Signal Processing}, vol.~28, no.~4, pp. 367--376, 1980.

\bibitem{fleuret2004fast}
F.~Fleuret, ``Fast binary feature selection with conditional mutual information.'' \emph{Journal of Machine learning research}, vol.~5, no.~9, 2004.

\bibitem{ispmi1}
F.~Zhang, W.~Zhou, Y.~Liu, W.~Geng, Y.~Shan, and C.~Zhang, ``Disentangling age and identity with a mutual information minimization for cross-age speaker verification,'' in \emph{Interspeech 2024}, 2024, pp. 3789--3793.

\bibitem{ispmi2}
J.~Li, J.~Han, S.~Deng, T.~Zheng, Y.~He, and G.~Zheng, ``Mutual information-based embedding decoupling for generalizable speaker verification,'' in \emph{Interspeech 2023}, 2023, pp. 3147--3151.

\bibitem{ispmi3}
S.~Yang, M.~Tantrawenith, H.~Zhuang, Z.~Wu, A.~Sun, J.~Wang, N.~Cheng, H.~Tang, X.~Zhao, J.~Wang, and H.~Meng, ``Speech representation disentanglement with adversarial mutual information learning for one-shot voice conversion,'' in \emph{Interspeech 2022}, 2022, pp. 2553--2557.

\bibitem{ispmi4}
W.~Kang, M.~J. Alam, and A.~Fathan, ``Mim-dg: Mutual information minimization-based domain generalization for speaker verification,'' in \emph{Interspeech 2022}, 2022, pp. 3674--3678.

\bibitem{ispmi5}
H.~Xue, X.~Wang, Y.~Zhang, L.~Xie, P.~Zhu, and M.~Bi, ``Learn2sing 2.0: Diffusion and mutual information-based target speaker svs by learning from singing teacher,'' in \emph{Interspeech 2022}, 2022, pp. 4267--4271.

\bibitem{ispfairness1}
P.~Dheram, M.~Ramakrishnan, A.~Raju, I.-F. Chen, B.~King, K.~Powell, M.~Saboowala, K.~Shetty, and A.~Stolcke, ``Toward fairness in speech recognition: Discovery and mitigation of performance disparities,'' in \emph{Interspeech 2022}, 2022, pp. 1268--1272.

\bibitem{ispfairness2}
I.-E. Veliche, Z.~Huang, V.~{Ayyat Kochaniyan}, F.~Peng, O.~Kalinli, and M.~L. Seltzer, ``Towards measuring fairness in speech recognition: Fair-speech dataset,'' in \emph{Interspeech 2024}, 2024, pp. 1385--1389.

\bibitem{ispfairness3}
W.-S. Chien and C.-C. Lee, ``An investigation of group versus individual fairness in perceptually fair speech emotion recognition,'' in \emph{Interspeech 2024}, 2024, pp. 3205--3209.

\bibitem{securespectra}
O.~Baser, K.~Kale, and S.~P. Chinchali, ``Securespectra: Safeguarding digital identity from deep fake threats via intelligent signatures,'' in \emph{Proc. Interspeech 2024}, 2024, pp. 1115--1119.

\bibitem{esfahanizadeh2023infoshape}
H.~Esfahanizadeh, W.~Wu, M.~Ghobadi, R.~Barzilay, and M.~M{\'e}dard, ``Infoshape: Task-based neural data shaping via mutual information,'' in \emph{ICASSP 2023-2023 IEEE International Conference on Acoustics, Speech and Signal Processing (ICASSP)}.\hskip 1em plus 0.5em minus 0.4em\relax IEEE, 2023, pp. 1--5.

\bibitem{kale2024texshape}
K.~Kale, H.~Esfahanizadeh, N.~Elias, O.~Baser, M.~M{\'e}dard, and S.~Vishwanath, ``Texshape: Information theoretic sentence embedding for language models,'' in \emph{2024 IEEE International Symposium on Information Theory (ISIT)}.\hskip 1em plus 0.5em minus 0.4em\relax IEEE, 2024, pp. 2038--2043.

\bibitem{donsker1975asymptotic}
M.~D. Donsker and S.~S. Varadhan, ``Asymptotic evaluation of certain markov process expectations for large time, i,'' \emph{Communications on pure and applied mathematics}, vol.~28, no.~1, pp. 1--47, 1975.

\bibitem{kraskov2004estimating}
A.~Kraskov, H.~St{\"o}gbauer, and P.~Grassberger, ``Estimating mutual information,'' \emph{Physical Review E—Statistical, Nonlinear, and Soft Matter Physics}, vol.~69, no.~6, p. 066138, 2004.

\bibitem{sgd}
H.~Robbins and S.~Monro, ``A stochastic approximation method,'' \emph{The annals of mathematical statistics}, pp. 400--407, 1951.

\bibitem{belghazi2018mine}
M.~I. Belghazi, A.~Baratin, S.~Rajeshwar, S.~Ozair, Y.~Bengio, A.~Courville, and D.~Hjelm, ``Mutual information neural estimation,'' in \emph{Proceedings of the 35th International Conference on Machine Learning}, vol.~80.\hskip 1em plus 0.5em minus 0.4em\relax PMLR, 10--15 Jul 2018, pp. 531--540.

\bibitem{songunderstanding}
J.~Song and S.~Ermon, ``Understanding the limitations of variational mutual information estimators,'' in \emph{International Conference on Learning Representations}.

\bibitem{hornik1989multilayer}
K.~Hornik, M.~Stinchcombe, and H.~White, ``Multilayer feedforward networks are universal approximators,'' \emph{Neural networks}, vol.~2, no.~5, pp. 359--366, 1989.

\bibitem{mozilladata}
R.~Ardila, M.~Branson, K.~Davis, M.~Henretty, M.~Kohler, J.~Meyer, R.~Morais, L.~Saunders, F.~M. Tyers, and G.~Weber, ``Common voice: A massively-multilingual speech corpus,'' in \emph{Proceedings of the 12th Conference on Language Resources and Evaluation (LREC 2020)}, 2020, pp. 4211--4215.

\bibitem{veaux2017cstr}
C.~Veaux, J.~Yamagishi, K.~MacDonald \emph{et~al.}, ``Cstr vctk corpus: English multi-speaker corpus for cstr voice cloning toolkit,'' \emph{University of Edinburgh. The Centre for Speech Technology Research (CSTR)}, vol.~6, p.~15, 2017.

\bibitem{dwork2006differential}
C.~Dwork, ``Differential privacy,'' in \emph{International colloquium on automata, languages, and programming}.\hskip 1em plus 0.5em minus 0.4em\relax Springer, 2006, pp. 1--12.

\bibitem{tsne}
L.~Van~der Maaten and G.~Hinton, ``Visualizing data using t-sne.'' \emph{Journal of machine learning research}, vol.~9, no.~11, 2008.

\bibitem{auroc}
J.~A. Hanley and B.~J. McNeil, ``The meaning and use of the area under a receiver operating characteristic (roc) curve.'' \emph{Radiology}, vol. 143, no.~1, pp. 29--36, 1982.

\end{thebibliography}

\end{document}